\RequirePackage{fix-cm}
\documentclass[twocolumn]{svjour3}          
\smartqed  
\usepackage{graphicx}
\usepackage{amsmath, amssymb}
\usepackage{multicol}
\usepackage[round]{natbib}
\usepackage{graphicx}
\usepackage{float}
\usepackage{subfig}
\usepackage{hyperref}
\usepackage[labelfont=bf]{caption}
\usepackage{titlesec}
\usepackage{microtype}
 \journalname{myjournal}
\begin{document}

\title{Laboratory model of electrovortex flow with thermal gradients, for liquid metal batteries
}


\author{J.S. Cheng      \and
        B. Wang         \and
        I. Mohammad     \and
        J.M. Forer        \and
        D.H. Kelley
}


\institute{J.S. Cheng \at
              Department of Mechanical Engineering \\
              University of Rochester \\
              Rochester, NY 14627 \\
              \email{j.s.cheng@rochester.edu}           
           \and
           D.~H.\ Kelley \at
              Department of Mechanical Engineering \\
              University of Rochester \\
              Rochester, NY 14627 \\
              \email{d.h.kelley@rochester.edu}
}

\date{Received: date / Accepted: date}

\maketitle

\begin{abstract}
We present a novel laboratory setup for studying the fluid dynamics in liquid metal batteries (LMBs). LMBs are a promising technology suited for grid-scale energy storage, but flows remain a confounding factor in determining their viability. Two important drivers of flow are thermal gradients, caused by internal heating during operation, and electrovortex flow (EVF), induced by diverging current densities. Our setup explores thermal gradients and electrovortex flow separately and in combination in a cylindrical layer of liquid gallium, simulating the behavior in a single layer of an LMB. In this work, we discuss the design principles underlying our choices of materials, thermal control, and current control. We also detail our diagnostic tools --- thermocouple measurements for temperature and Ultrasonic Doppler Velocimetry (UDV) probes for velocities --- and the design principles which go into choosing their placement on the setup. We also include a discussion of our post-processing tools for quantifying and visualizing the flow. Finally, we validate convection and EVF in our setup: we show that scaling relationships between the nondimensional parameters produced by our data agree well with theory and previous studies.
\end{abstract}

\section{Introduction}

The electricity generated by renewable energy sources almost doubled over the past decade in the United States \citep{USelecSources} and is likely to continue to rise in the near future. However, the intermittent nature of renewable sources such as solar and wind presents a challenge for widespread integration into the electrical grid: the demand for electrical power can vary by up to 50\% in a given region of the US over the course of a day \citep{Hodge20}, but this variation rarely lines up favorably with peak renewable production. To facilitate the adoption of renewable sources, grid scale energy storage solutions are needed. The liquid metal battery (LMB) is one promising technology due to its low cost and long life \citep{Wang14,Perez15}.

LMBs are electrochemical cells that consist of two liquid metal electrodes separated by a layer of molten salt electrolyte, and with all three layers stably stratified. The more electropositive metal, an electron donor, is the negative electrode (anode), while the more electronegative metal, an electron acceptor, is the positive electrode (cathode). The interaction between the two electrodes creates a thermodynamical cell voltage. During discharge, the negative electrode metal is oxidized into cations and electrons. The molten salt electrolyte only allows cations through to the positive electrode metal forcing the electrons through an external circuit thus providing electrical power \citep{Kim13}. The reverse happens during charging where electrical power is absorbed.  Much of the cost effectiveness of LMBs is owed to the liquid nature of the battery: the layers naturally separate by density which leads to simpler construction and improved ability to scale to larger or smaller cells, the liquid electrodes are immune to the microscale damage suffered by solid electrodes during charge and discharge, and the candidate materials can be relatively inexpensive \citep{Bradwell12, Kim13,Wang14,Li16,Dai18}. LMBs are a developing technology, with various container geometries and chemical compositions under exploration \citep{Bradwell12, Poizeau12, Kim13jps, Ning15, Gong20}.

%
\begin{figure*}
\centering
\includegraphics[width=\textwidth]{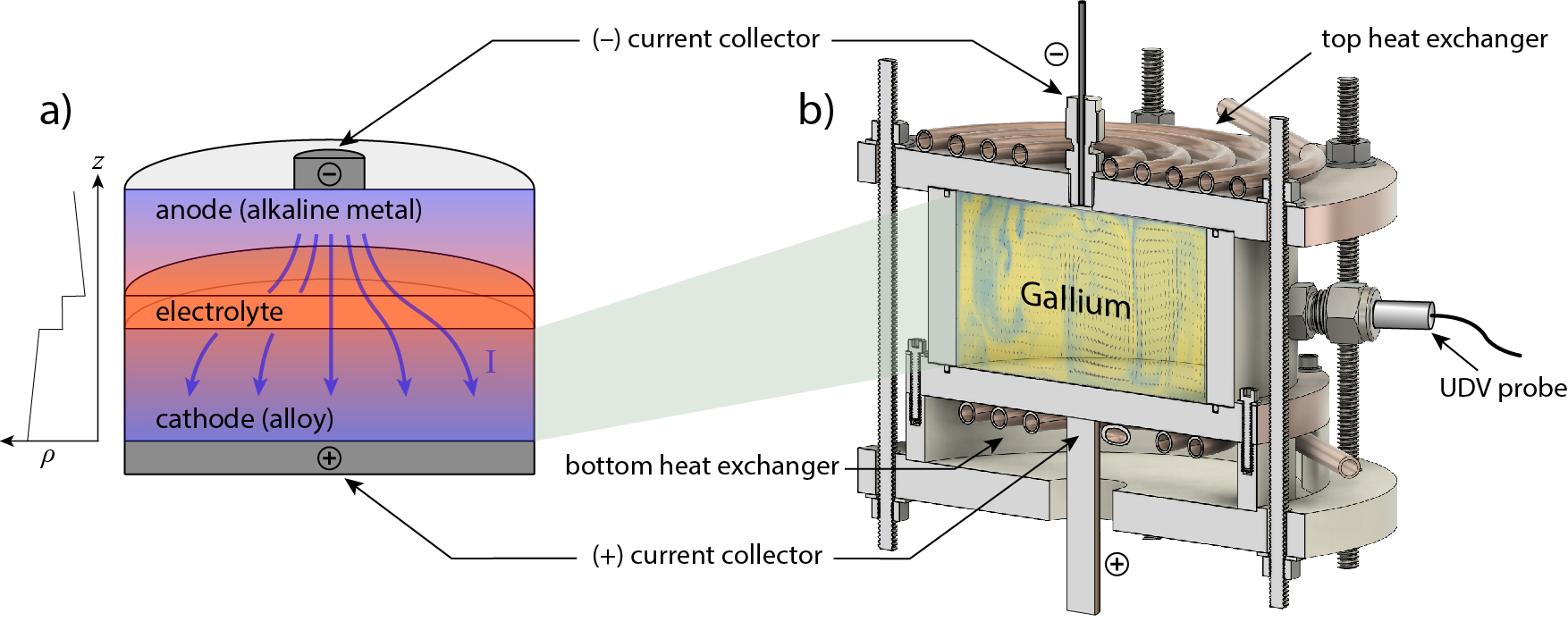}
\caption{\label{F:schem}(a) Schematic demonstrating thermal gradients and electromagnetic forces in a liquid metal battery and (b) our experimental setup for studying these forces and their interactions. The diagram in panel (a) shows the variation in density $\rho$ with vertical position $z$, demonstrating that the cathode, electrolyte, and anode layers are separated by stable stratification. Each layer also contains density gradients due to heating in the electrolyte layer. The experimental setup can model either the cathode or anode of the battery, depending on whether we impose a stable or unstable temperature gradient. The velocity field shown in panel (b) is for illustration only and was adapted from \cite{Personnettaz19}.}
\end{figure*}

While most earlier analyses focused on the electrochemistry and economic viability, recent years have brought attention to the fluid dynamics occurring within the battery during operation \citep{Kelley18,Weber18}. The fluid nature of LMBs lends them a strong propensity for flow: a variety of thermal, chemical, and electromagnetic forces play against one another with serious consequences, both helpful and harmful, for the battery. Helpfully, flow can prevent buildup of intermetallics, a primary source of failure in LMB test cells. During discharge, flow redistributes ion-saturated parcels of cathode material away from the interface faster than diffusion and can thus improve efficiency. Harmfully, sufficiently vigorous flows can disturb the interfaces between different fluid layers and cause a short, destroying the battery. The fluid dynamics in LMBs can be paramount to their operation and efficiency. It is thus important to develop a broad understanding of how various flow drivers interact with one another via scaling laws between governing parameters that can be generalized.

Here, we present a novel laboratory device for exploring the interaction between two prominent drivers of fluid motion in LMBs -- thermal gradients and electro-vortex flow (EVF) -- over broad ranges of the governing parameters. By covering a large swath of parameter space, we can construct scaling laws and regime diagrams to extrapolate results to real battery systems. We apply these forces to a layer of liquid gallium to simulate behaviors in a single LMB layer, collecting temperature data using thermocouples and velocity data using Ultrasonic Doppler Velocimetry (UDV) probes. 

In this paper we will discuss the design principles going into the construction of the experimental setup, laboratory techniques for optimal acquisition of data, and novel post-processing techniques for interpreting experimental results. In section \ref{sec:motiv}, we describe the framework of flow physics we intend to capture with our device and how these lead to our overarching design choices. In section \ref{sec:design}, we delve into detail about the design, explaining the specifics of how we control the flow with precision and ensure the setup is robust. In section \ref{sec:diag}, we describe the expected behavior under the imposed flow drivers and how our diagnostics are arranged to perform meaningful measurements.
In section \ref{sec:data}, we validate our design by showing early data and comparison with predictions from the literature.
In section \ref{sec:conclusion}, we provide concluding remarks.

\begin{table*}
\caption{Estimated material properties of liquid gallium at 43$^\circ$C \citep{Okada92,Brito01,Aurnou18} compared to ranges of properties for LMB candidate materials at their melting temperatures \citep{Kelley18}. These properties are comparable, particularly with the cathode, though it should be noted that broad ranges of values are possible in LMBs. Anode materials considered are Li, Mg, and Na \citep{Davidson68,Iida15,Iida15b}
while cathode materials considered are Bi, Pb, Sb, and Zn \citep{Iida15,Iida15b,Fazio15}.
Properties listed are mass density $\rho$, kinematic viscosity $\nu$, thermal diffusivity $\kappa$, volumetric expansion coefficient $\alpha$, Prandtl number $Pr$, electrical conductivity $\sigma$, and speed of sound $c$.}
\label{tab:Ga}
\begin{tabular*}{\textwidth}{l @{\extracolsep{\fill}} lllllll}
\hline\noalign{\smallskip}
Material & $\rho$ [kg/m\textsuperscript{3}] & $\nu$ [m\textsuperscript{2}/s] & $\kappa$ [m\textsuperscript{2}/s] & $\alpha$ [1/K] & $Pr$ & $\sigma$ [S/m] \\
\noalign{\smallskip}\hline\noalign{\smallskip}
Gallium & 5870 & $3.588\times10^{-7}$ & $1.345\times10^{-5}$ & $1.25\times10^{-4}$ & 0.027 & $3.88\times10^{6}$ \\
Anode & 500--1600 & 7.5--12$\times10^{-7}$ & 2--7$\times10^{-5}$ & 1.6--2.5$\times10^{-4}$ & 0.01--0.06 & 3.6--10$\times10^{6}$ \\
Cathode & 6500--10000 & 1.6--5.3$\times10^{-7}$ & 1.0--1.6$\times10^{-5}$ & 1.2--1.5$\times10^{-4}$ &  0.01--0.03  & 0.8--2.7$\times10^{6}$ \\
\noalign{\smallskip}\hline
\end{tabular*}
\end{table*}

\section{Design Motivation} \label{sec:motiv}		

The LMB environment is host to many flow drivers and interactions. We choose to focus on two prominent contributors to fluid motion: thermal gradients and electrovortex flow \citep{Bojarevics89,Davidson01,Kelley14,ShenZikanov16,Personnettaz18,Weber20,Keogh21}. Figure \ref{F:schem}a is a schematic of a liquid metal battery during operation, demonstrating how each flow manifests. Thermal gradients across the layer cause stable stratification or convection, while a diverging current density across the layer causes electrovortex flow. We describe the physics behind each flow below in sections \ref{sec:grad} and \ref{sec:evf}.

Our experimental setup --- shown in Figure \ref{F:schem}b --- simulates the behavior in the cathode or the anode of the LMB by producing thermal gradients and electrovortex flow together in an enclosed fluid layer. We focus on each layer independently, precluding interfacial flows that make it difficult to pinpoint the source of a given observed motion~\citep{Kollner17}. We chose liquid gallium as the primary working fluid: it offers similar properties to LMB candidate materials, but near room temperature, rather than at hundreds of degrees Celcius (see Table \ref{tab:Ga}). Of particular relevance to the fluid dynamics is the Prandtl number $Pr = \nu/\kappa$, where $\nu$ is the kinematic diffusivity and $\kappa$ is the thermal diffusivity, which tends to have values between 0.01 and 0.03 for LMB layers and has value 0.027 in our experimental setup. Operating around room temperature affords us a greater degree of thermal control, since water cooling and heating become viable (discussed further in Section \ref{sec:contr}).

We study cylindrical LMBs to avoid introducing another free parameter in the aspect ratio between length versus width. Cylinders are also the most common shape for models in the literature \citep{Ning15,Li16,Beltran17,Weber18}. Our cell diameter $D = 10$~cm is chosen to achieve relevant values of the nondimensional parameters (detailed below in sections \ref{sec:grad} and \ref{sec:evf}) and ensuring scalability to LMBs while also keeping experiments manageable in time and length scale. The depths of the cathode and anode layers change during battery operation, as material passes from one layer to the other. Different cylindrical aspect ratios $\Gamma = D/H$ (where $H$ is height) can contain significantly different flows \citep{Wagner13,Vogt18,Wang20}, and these flows may vary in their effect on LMB mixing. We employ interchangeable side wall segments ranging from height $H=3.3-7.1$~cm ($\Gamma = D/H = \sqrt{2}-3$) to address this.

\subsection{Thermal gradients} \label{sec:grad}
In liquid metal batteries, thermal gradients occur because of the high electrical resistivity of the electrolyte layer --- typically several orders of magnitude more resistive than the other layers \citep{Janz68}. The electrolyte preferentially heats up as cations flow through it during both charge and discharge of the battery. The anode experiences a destabilizing thermal gradient that leads to convection, whose forcing strength is characterized by the nondimensional Rayleigh number 
\begin{equation}
\label{eq:Ra}
Ra = \frac{\alpha g \Delta T H^3}{\nu\kappa} \, .
\end{equation}
Here, $\alpha$ is the coefficient of thermal expansion, $g$ is gravitational acceleration, and $\Delta T$ is the magnitude of the temperature difference. Beyond a critical value $Ra \sim 2000$ \citep{Chandra61} and prior to $Ra \sim 10^9$ convective flows remain within a regime of dynamical similarity \citep{Cioni97,Ahlers09}. Liberal estimates for typical LMB temperature gradients and dimensions place them squarely within this regime:
even for an upper bound LMB size of 50 cm, temperature gradients upwards of $\sim 100~^\circ$C would be required to reach a higher regime of convection. Conversely, even for a lower bound LMB size of 5 cm, a temperature gradients of order 0.1~$^\circ$C is enough to reach the onset of the convection regime we explore. With this in mind, our setup is designed to access $Ra$ values spanning two orders of magnitude, from $10^4$ to $10^6$.
Since $Ra$ scales linearly with $\Delta T$, we need detailed temperature measurements to characterize $Ra$. We discuss these in Section \ref{sec:diag}.

One way to quantify the resultant flows is with the Reynolds number $Re=UL/\nu$, where $L$ is a typical length scale and $U$ is a typical flow speed. The Reynolds number serves as an overall descriptor of system dynamics, with literature predictions for its scaling in both convection and EVF. In convection, $Re$ is expected to follow a power law against $Ra$~\citep{Ahlers09}:
\begin{equation}
\label{eq:rera}
Re \propto Ra^{2/5}Pr^{-3/5} \, ,
\end{equation}
for the parameter ranges relevant to LMBs. Velocities therefore scale with the temperature gradient, and $Ra$ can provide insight on whether the flow is in the range of helpful mixing or strong enough to disrupt fluid interfaces.

We use this to validate our setup in section \ref{sec:data}. However, the Reynolds number alone does not tell us the morphology of the flow, which can vary depending on flow forcing and geometry of the system. Velocity measurements that capture the flow from various angles and across different spatial positions are therefore a priority. More details about our velocity diagnostics are discussed in section \ref{sec:diag}.

We note that recent studies estimate that compositional gradients will dominate thermal gradients in the cathode during charge and discharge \citep{Personnettaz19,Herreman20}. The nondimensional parameters for solutal convection are structurally similar to thermal convection, with a solutal Prandtl number $Pr_s = \nu/\mathfrak{d}$, analogous to the thermal Prandtl number $Pr$ ($\mathfrak{d}$ is the mass diffusion coefficient). This similarity means our experimental setup may also provide insight into compositional gradients when water ($Pr\simeq7$) is the working fluid instead of gallium. 

\subsection{Electrovortex flow} \label{sec:evf}
Electrovortex flow occurs when a current density diverges radially \citep{Shercliff70}, as is the case in almost any liquid metal battery configuration (shown by the blue arrows in Figure \ref{F:schem}a). The induced magnetic field from a given current line interacts with nearby currents and generates Lorentz forces which drive flow radially toward the smaller current collector. The effect strengthens closer to the current collector, resulting in a downward-pointing jet under the current collector, and induced flow throughout the rest of the fluid layer. The strength of EVF is characterized by a so-called EVF parameter \citep{Lundquist69}:
\begin{equation}
\label{eq:S}
S = \frac{\mu_0 I^2}{4 \pi^2 \rho\nu^2}
\end{equation}
where $\mu_0$ is the vacuum permeability and $I$ is the imposed current. 
An especially vigorous jet has potential to disrupt the interfaces between LMB layers. A well-documented transition toward an azimuthal swirling flow also occurs \citep{Bojarevics89,Davidson01}, which we describe in more detail in section \ref{sec:diag}. The swirling flow may be more well-suited to mixing within the cathode with less risk of interface disruption \citep{Weber20}. \cite{Ashour18} observed this transition at a current of 20 A or $S = 1.7 \times 10^4$ in an experiment using PbBi eutectic. Simulations by \cite{Herreman19} find different velocity scalings on either side of this transition, with $Re \sim S$ for $S \lesssim 10^3$ and $Re \sim S^{1/2}$ for $S \gtrsim 10^5$. To ensure we can explore the transition, our setup is designed to access $S$ values from 0 to $3\times10^5$.

The combination of EVF and convection has been studied in the past. Kelley and Sadoway \cite{Kelley14} found that introducing EVF in a convecting fluid causes the flow to become more ordered even at small input currents, an observation corroborated by \cite{Ashour18} and \cite{Keogh21}. Predictions also exist for the transition between convection-dominated and EVF-dominated flow:  \cite{Davidson01} derived a criterion where convection begins to dominate when $\left(RaPr\right)^{3/7} \gtrsim S^{1/2}Pr$, although this is for a hemispherical geometry rather than cylindrical. \cite{Ashour18} argued that a transition occurs when the characteristic velocities for each forcing become comparable. Our accessible $Ra$ and $S$ ranges ensure that we can test these predictions over broad parameter ranges.

During LMB operation, a stabilizing thermal gradient manifests in the cathode layer that may suppress other sources of fluid motion. To our knowledge, stable thermal stratification combined with EVF has not been previously explored although \cite{Herreman20} examine EVF with solutal stratification. We therefore make it possible to reverse the thermal gradient in our setup to study this problem. Data produced by our apparatus will be a source of novel information about the LMB cathode.

\begin{figure}
\centering
\includegraphics[width=\linewidth]{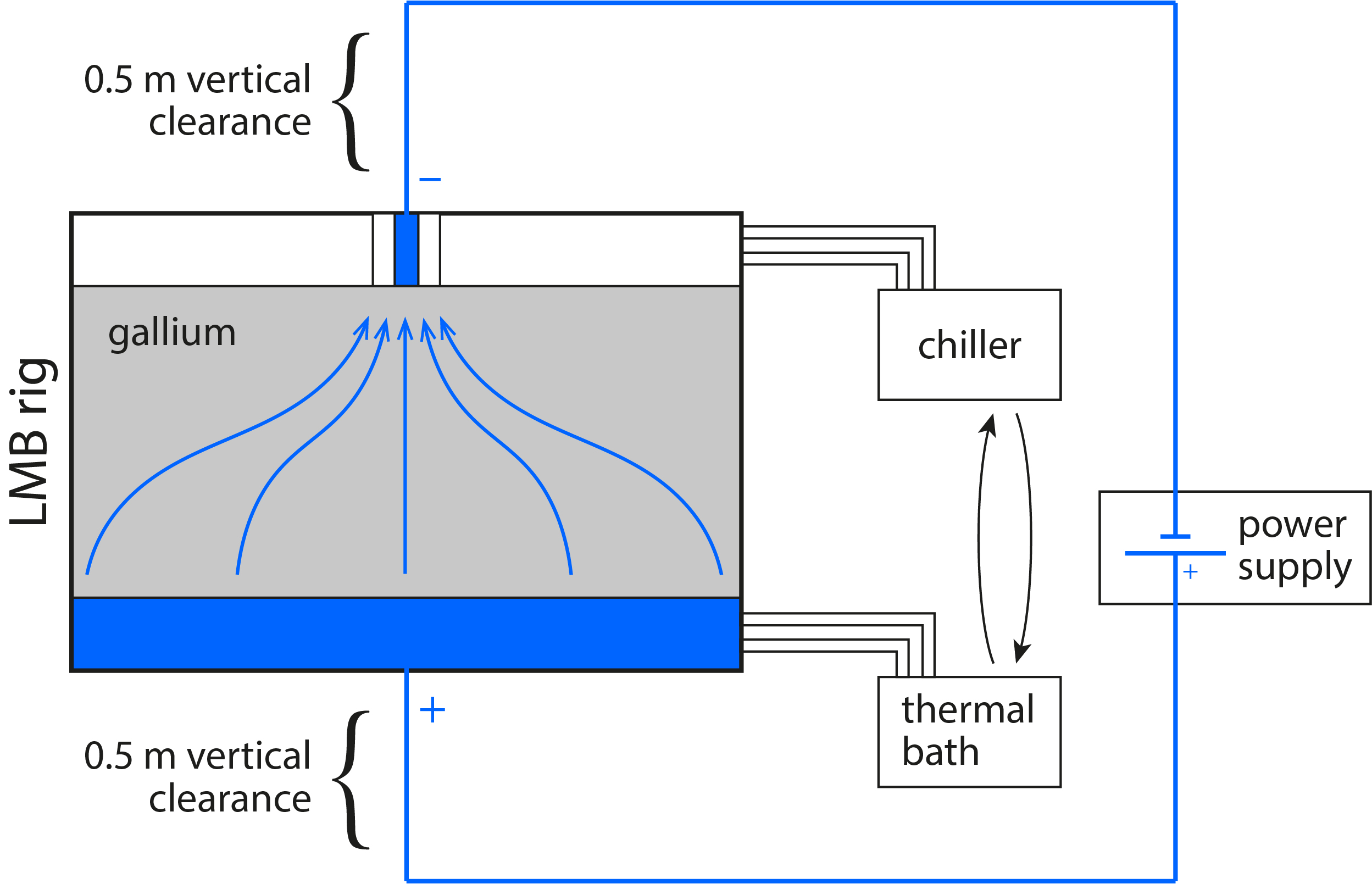}
\caption{\label{F:circ}Current and thermal gradient controls for the experimental setup. Heat exchangers are composed of a copper plate and copper coil. Water is recirculated through the coils by a thermal bath and a chiller, which can be interchanged to induce stable or unstable stratification. The current pathway is drawn in blue. The bottom thermal block also acts a current collector. The wires connected to the current collectors are routed vertically for a large distance such that the magnetic field induced by the current does not interfere with the flow.}
\end{figure}

\section{Experimental design} \label{sec:design}

\subsection{Controls} \label{sec:contr}
Figure \ref{F:circ} is a diagram of the experimental setup and its associated external components for controlling temperature and current. The gallium layer is bounded above and below by copper plates, each plate soldered to a copper coil which serves as a heat exchanger: two thermal baths recirculate water through each of these coils such that either a stabilizing or destabilizing temperature gradient can be imposed on the gallium layer. The coils are wound in a non-inductive double spiral pattern to minimize horizontal temperature gradients while accommodating instruments and the negative electrode. 

%
\begin{figure*}
\centering
\includegraphics[width=0.8\textwidth]{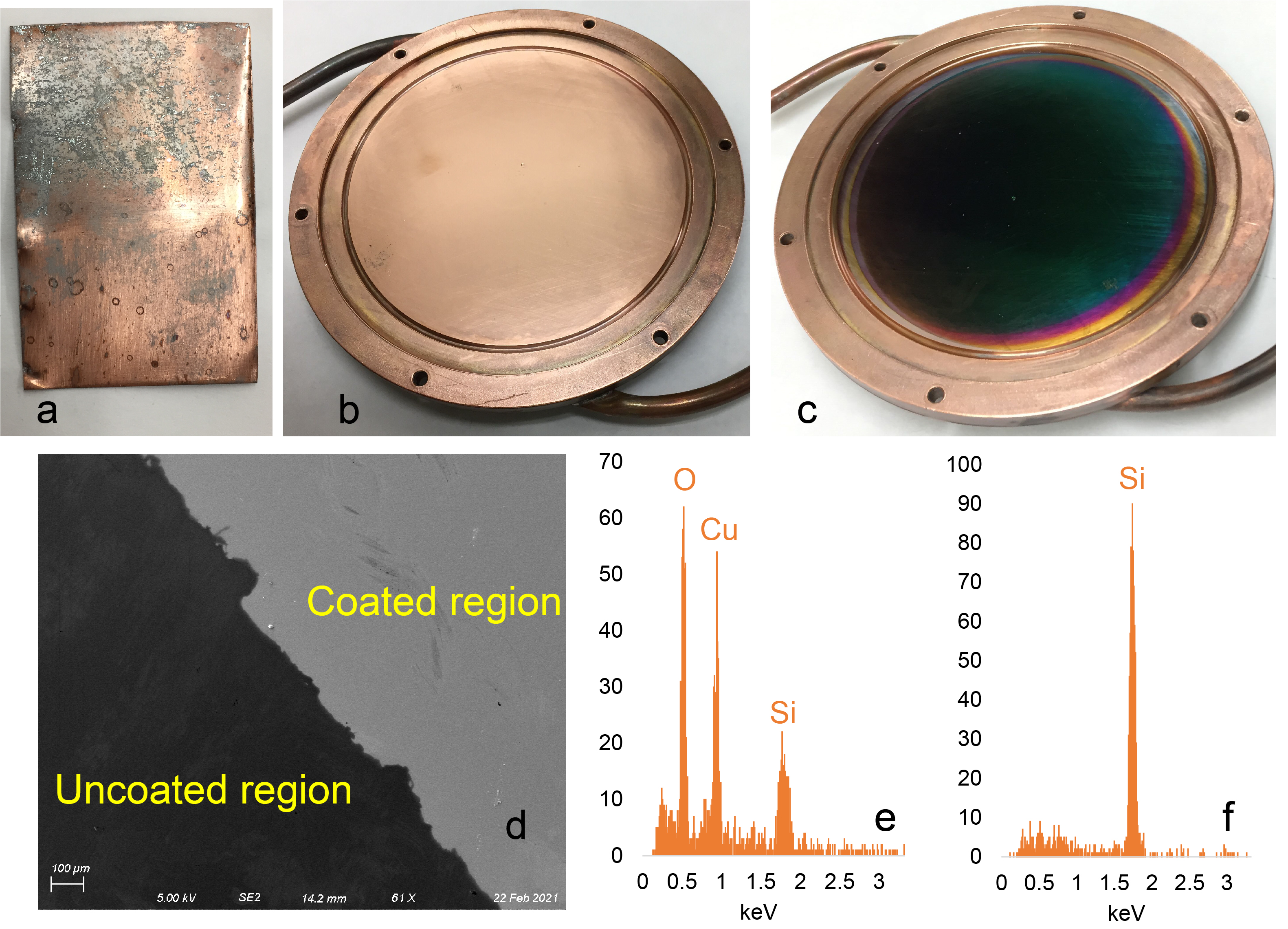}
\caption{\label{F:coat}Coating the thermal blocks where they contact gallium. (a), A (3 cm $\times$ 5 cm) sheet of copper corroded by contact with liquid gallium over the course of two days. (b), Before CuO coating. (c), After CuO coating. (d), The boundary between coated and uncoated regions on a piece of Si, coated using the same procedure and imaged via electron microscopy, which showed good coverage. (e–f), Energy-dispersive x-ray spectroscopy indicated nearly pure Si in the uncoated region and CuO in the coated region.
}
\end{figure*}
%

Copper is used for the heat exchangers due to its high thermal and electrical conductivity. However, liquid gallium attacks and damages copper, as shown in Figure \ref{F:coat}a. Such corrosion can be counteracted by a protective coating; following \cite{ZhangH07}, we choose copper oxide for the coating layer. 
A thermal evaporation method \citep{Ozer93} was performed using a Ladd Evaporator (Ladd Research Industries, U.S.A). During the evaporation process, copper oxide pellets (CuO, 99.9\% purity) were placed in a tungsten basket and evaporated in the vacuum chamber under a base pressure of $5\times10^{-4}$ Pa, depositing copper oxide particles on the heat exchanger suspended above. The resulting coating is shown in Figure \ref{F:coat}c, and the X-ray spectrum in Figure \ref{F:coat}d demonstrates that the process was successful. \textit{In situ} experiments confirm that this layer prevents gallium corrosion. Based on the evaporation rate, we estimate the thickness of the coating layer as 0.6~{\textmu}m. A 1-D Fourier's law approximation (assuming the copper plate and coating layer as resistors in series) finds the layer's effect on heat transfer to be negligible, with only a 0.15\% change in $\Delta T$ compared to a pure copper heat exchanger.

To ensure temperature uniformity across each plate, we must account for the impacting and detaching of convective plumes on the surface: these should not create significant temperature gradients within the plate. We address this concern by considering an estimate of the Biot number for convection
\begin{equation}
\label{eq:Bi}
Bi = \frac{q}{\Delta T} \frac{h_\mathrm{plate}}{k_\mathrm{plate}} = \frac{k}{k_\mathrm{plate}}\frac{h_\mathrm{plate}}{H}Nu,
\end{equation}
where $h_\mathrm{plate}$ is the thickness of the plate, $k_\mathrm{plate}$ is the thermal conductivity of the plate, $k$ is the thermal conductivity of gallium, and the Nusselt number $Nu$ is a nondimensional parameter representing the convective heat transfer efficiency. Typically, $Bi < 0.1$ means that the temperature inside the plate remains uniform. Assuming that $Nu$ scales with $Ra$ as shown in the literature \citep{Ahlers09}, we estimate that a plate thickness $h_\mathrm{plate} < 0.8$ cm would give uniform temperature.

We must also ensure that the temperature at the surface in contact with the gallium layer is insensitive to the structure of the cooling coil. We address this by creating a simplified, two-dimensional model of heat transfer --- considering a cross-section of the copper plate with the coil as a series of hot and cold patches --- and finding the steady-state temperature distribution at the plate-gallium interface. See the Appendix for a more thorough description. We found that when the plate thickness is 0.5 cm, in all cases the temperature deviation is $<2$\%. Ultimately, 0.5 cm was chosen as the plate thickness to minimize $Bi$, maximize temperature uniformity, and also optimize machining efficiency.

The side wall of the vessel is a 1.1~cm thick hollow cylinder made of Delrin (polyoxymethylene), a thermally resistive material ($k_\mathrm{Delrin} = 0.37$ W/mK). This ensures that heat flows almost exclusively through the gallium layer: using a 1-D Fourier's law approximation (assuming the Delrin sidewall and gallium layer as resistors in parallel) yields that less than 0.6\% of the applied heat flux travels through the sidewalls for all experiments (and is a notable overestimate because convection has not been taken into account).  

The entire experimental setup is surrounded by a layer of foam rubber insulation that is at least 1 inch thick to minimize heat losses, ensuring that nearly all the heat introduced by the thermal bath is extracted by the chiller, and maximizing the range of achievable temperature differences $\Delta T$.

\begin{figure}
\centering
\includegraphics[width=1\linewidth]{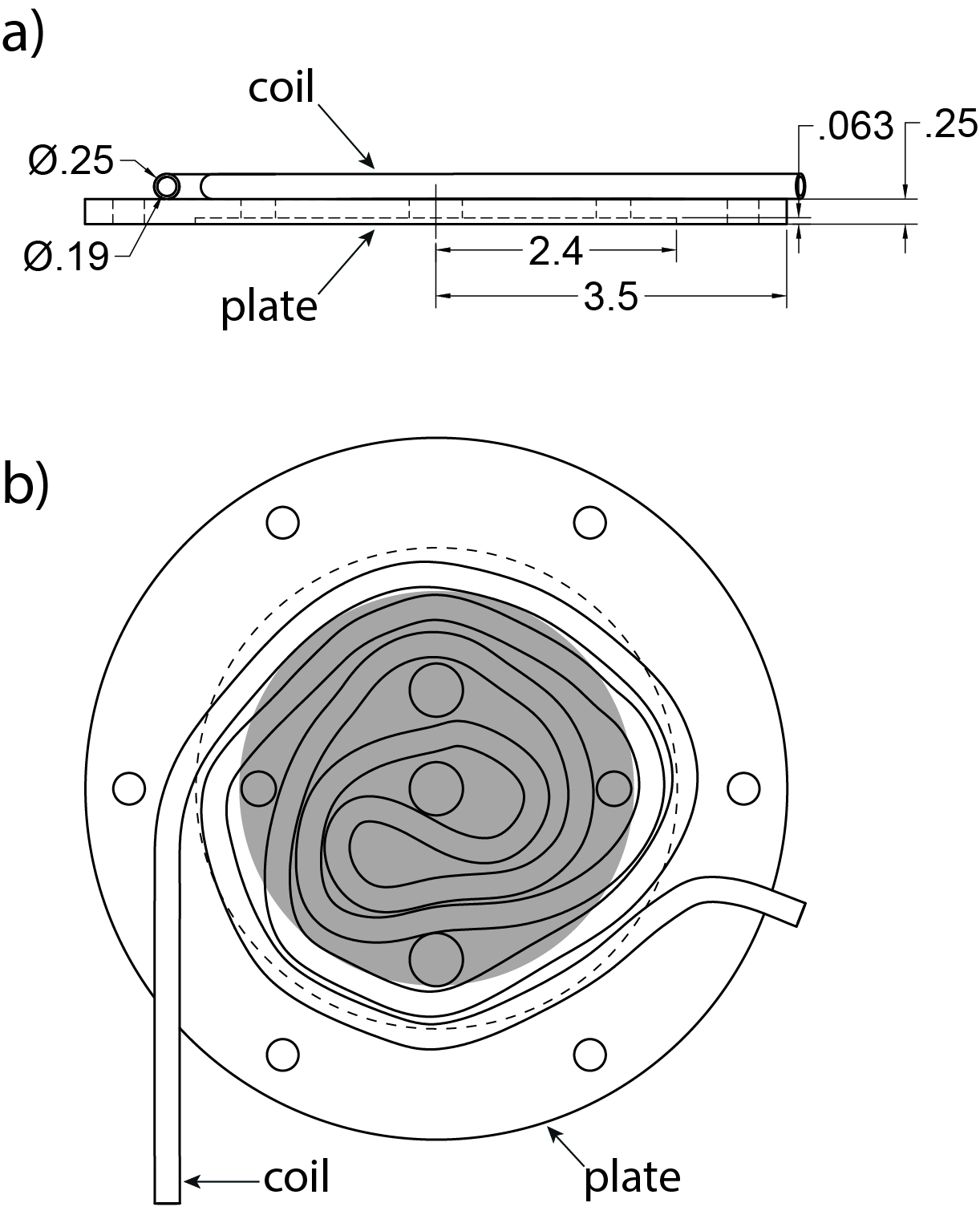}
\caption{\label{F:coil} The copper coil and plate which form the top thermal block, seen from the side (a) and from the top (b). Dimensions are given in inches. We minimize spacing between coils and ensure non-inductive winding while accommodating the UDV probes, filling ports, and top electrode. The dashed line marks the outside of the vessel side wall while the gray region indicates the position of the gallium layer.}
\end{figure}

We induce currents of 0 to 90~A through the system with a TDK-Lambda power supply. Delrin's low electrical conductivity (0.2 W/mK) ensures all currents pass through the gallium layer instead of traveling along the side wall. The negative electrode is a copper rod of multiple possible diameters (0.318 cm, 0.476 cm, 0.635 cm) attached to the top plate by a plastic fitting, electrically isolating it from the top plate. The bottom plate is mated to a copper rod connected to the power supply: with the low electrical resistivity of copper --- roughly a factor of 15 lower than liquid gallium \citep{Ginter86} --- the current spreads out uniformly and the entire bottom plate forms the positive electrode. The large diameter ratio between electrodes ensures a large horizontal divergence in the current paths to trigger EVF. 

Electrovortex flow is sensitive to external magnetic fields \citep{Davidson01}. To minimize the induced field from horizontal currents at the gallium layer, we ensure that the currents flowing in and out of the setup are oriented only vertically in the vicinity of the gallium: the vertical rods attached to the top and bottom plates for carrying current are at least 0.5 m long (see Fig. \ref{F:circ}). Using the Biot-Savart law, we estimate that the induced lateral magnetic field strength is $< 20$~\textmu T at the fluid layer (smaller than Earth's magnetic field).

\section{Diagnostics} \label{sec:diag}

Temperature data are collected by K-type thermocouple probes placed on the top and bottom heat exchangers. They are distributed around the plates at various radial distances and separated by $90^\circ$ in angle (find distances). Small blind holes are drilled into the thermal plates to assign reproducible probe locations and to bring the tips of the probes into closer vicinity of the fluid layer (within 3 mm). Signals are digitized and recorded by National Instruments NI 9211 cDAQ thermocouple input modules, operated by a LabVIEW acquisition program. Temperature measurements show that the spread in temperatures between probes on the same plate is typically less than 10\% of the temperature difference across the gallium layer in all cases, and usually close to 5\%. Furthermore, the spread in  temperatures on a given plate is only double the standard deviation of an individual thermocouple probe. We believe this demonstrates a successful degree of temperature uniformity over each heat exchanger.

Liquid gallium is opaque, which greatly limits our velocity measurement options: particle tracking, Schlieren, and dye based techniques are not viable. UDV is therefore our method of choice. In this technique, pulsed ultrasound waves are emitted by transducers, fluid provide an instantaneous velocity profile along the sound propagation direction. This technique is non-invasive and has been validated for use with various liquid metals \citep{Takeda87,Brito01,Eckert07}. 

For gallium experiments, we do not introduce any tracer particles. Naturally existing gallium oxide inclusions are capable of scattering the ultrasound waves. Due to the density difference between liquid gallium and gallium oxide, though, scattering particles tend to settle over time, reducing scattering strength. The wetting and acoustic coupling conditions between the UDV transducer and gallium are also observed to deteriorate over time \citep{WangKelley21}. We therefore limit the duration of each set of UDV measurements to 1 hour for optimal signal quality. 

\begin{figure}
\centering
\includegraphics[width=0.8\linewidth]{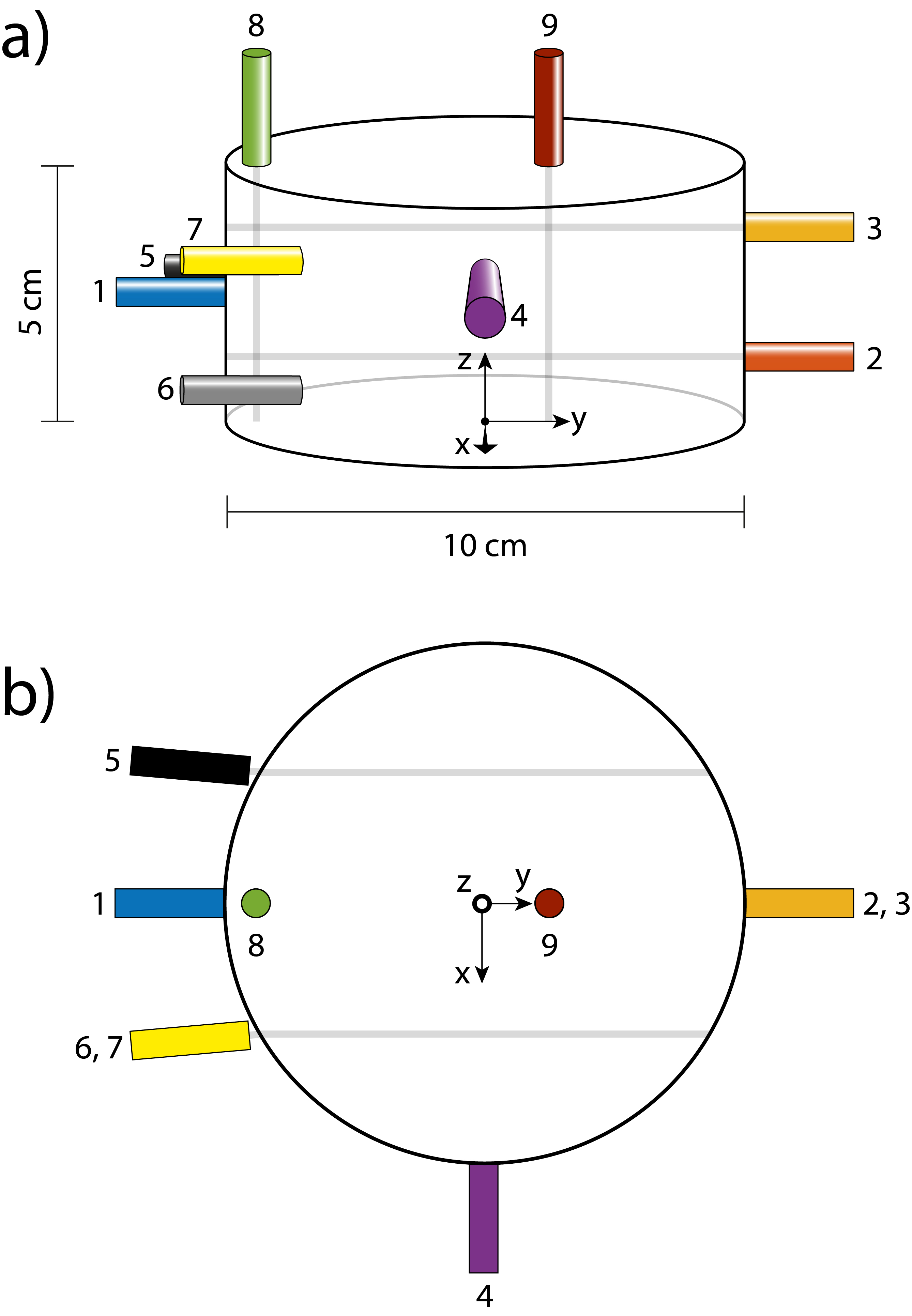}
\caption{\label{F:probes} UDV probe locations on the 50 mm high setup (other tank heights use analogous locations), seen from the side (a) and from above (b). Each probe measures one component of velocity along the probe beam. 
Probes 5,6 and 7 are placed at chord positions and angled 5$^\circ$ away from the y-direction, such that their beams are oriented in the y-direction after passing through the sidewall (ie. parallel to probes 1,2 and 3).
}
\end{figure}

The working principle of UDV has been introduced in detail in our previous work \citep{Perez15}. Here, we use nine ultrasound transducers (Signal Processing, Switzerland) to measure flow velocities, each of which measure the component of velocity towards and away from the probe. Placements and orientation are shown in Figure \ref{F:probes}. Because of the sparse, `gappy' nature of UDV data, we must carefully choose probe locations to best capture the expected flows. We also have to take into account the geometric limitations imposed by the setup: each probe is held in place by a compression fitting of finite size (SwageLok 8~mm), limiting the density with which probes can be arranged. For probes in direct contact with the gallium, the fitting seals around the probe and provides a flush surface with the inner walls of the vessel, but also requires extra clearance.
Our UDV measurements are validated via comparison to particle tracking velocimetry measurements --- see \cite{WangKelley21} for details. 

%

Convection and EVF are both expected to take the form of overturning rolls in the accessible parameter ranges and geometries \citep{Cioni97,Davidson01}. In convection, a single convection roll is expected to fill the geometry whereas in EVF, a vertical jet descends along the centerline inducing rising flow along the side walls \citep{Davidson01,Ashour18}. We therefore place two transducers with working frequency 8 MHz (probes 8 and 9) at the top lid and in direct contact with gallium for measuring vertical velocities at different radial positions. One is 4 mm from the side wall ($x=0.92D/2$) and the other is at the center or 12.5 mm from the center ($x=0.25D/2$), depending on the case (the center position is occupied by the negative electrode during EVF cases, but is available during convection-only cases).
\begin{figure*}
\centering
\includegraphics[width=1\textwidth]{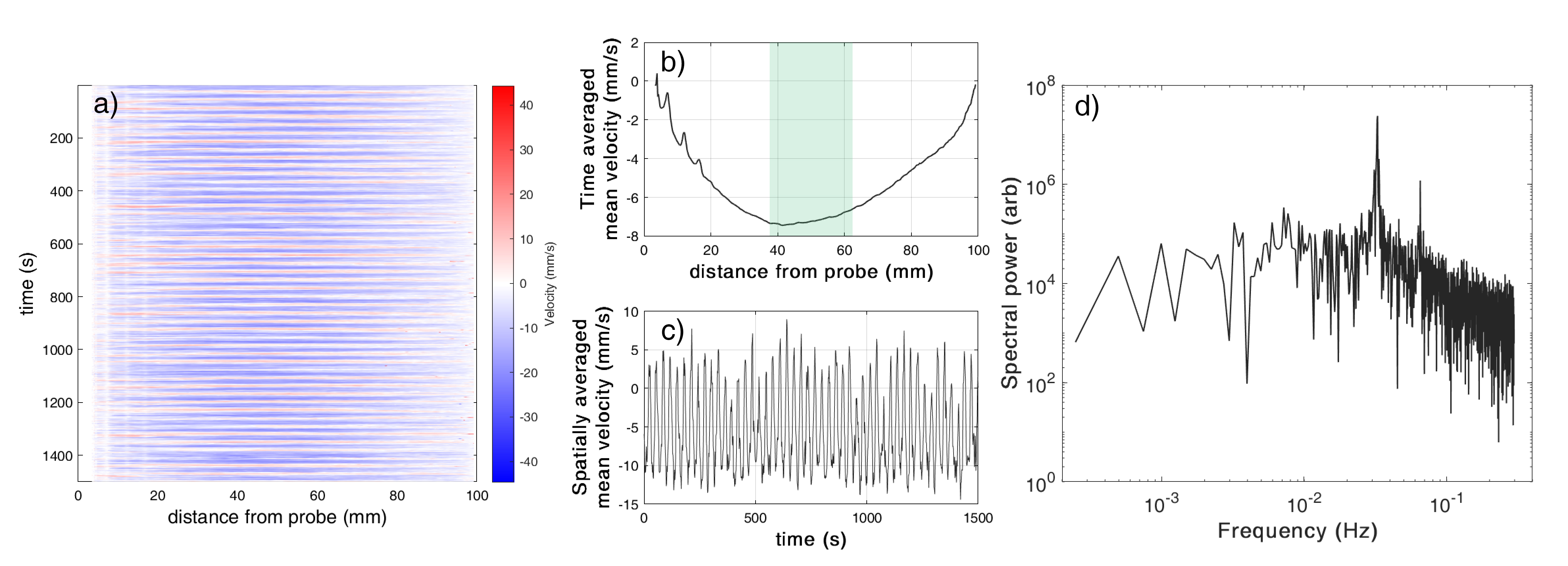}
\caption{\label{F:data}UDV measurements of convection at $Ra=3.7\times10^5$ in the $\Gamma=1.7$ setup. (a), Velocity measured by probe 2 (see Fig. \ref{F:probes}), varying over space and time. (b), Time-averaged velocity varying with distance from probe 2. The green highlighted region is used for $Re$ calculations to produce Figure \ref{F:rera}. (c), Spatially averaged mean velocity varying over time. (d), The corresponding power spectrum has a prominent peak at 0.033 Hz.}
\end{figure*}

Four transducers with working frequency 8 MHz are placed at the side wall of the vessel, oriented in the radial direction and in direct contact with the gallium. Convection rolls are likely to span the whole vessel in the vertical direction in our $\Gamma > 1$ geometries, and strong horizontal flows should be situated near the top and bottom of the tank. In addition, horizontal velocities are expected to show vertical asymmetry when stable stratification is introduced \citep{Herreman20}. We therefore place horizontal probes at $z=0.75H$ and $z=0.25H$. Since flows are not necessarily axisymmetric \citep{Vogt18,Keogh21}, we place probe 4 on the x-z plane instead of the y-z plane. 

Flows with strong azimuthal components, such as the swirl flow in EVF, cannot be detected by radial probes. We therefore use three horizontal probes oriented along chord positions $\left(x=-0.25D, 0.25D\right)$ and at heights $z=\left(0.25H, 0.5H, 0.75H\right)$. These transducers (probes 5 through 7) have a working frequency of 4 MHz and measure flow velocities through the wall of the vessel (no direct contact with gallium). An 8~MHz ultrasound beam has a smaller wavelength and thus gives better spatial resolution. A 4~MHz ultrasound beam, on the other hand, has a lower attenuation rate and thus allows stronger signals to be transmitted through the vessel wall. To maximize the transmission of ultrasound signal, the area of the side wall where transducers make contact is thinned to 0.3 mm.

The speed of sound in Delrin (2430 m/s) is slower than in gallium (2860 m/s), and refraction occurs when the transducer beam passes into the fluid layer. In order to keep the chord probe beams parallel to probes 1--3, we used Snell's law to determine that the probes should be tilted 5$^\circ$ away from the y-direction (see Fig.~\ref{F:probes}b).

All transducers are connected to a DOP 3010 Velocimeter (Signal Processing, Switzerland) and operated in emit/receive mode for data acquisition. The Pulse Repetition Frequency (PRF) and number of emissions per profile together determine the sampling frequency and measurement quality of UDV. During measurements, those two parameters are adjusted according to the flow conditions -- for example, it is necessary to ensure that the sampling frequency is always faster than the variation of the flow velocities. To compensate for ultrasound attenuation in gallium, the Time Gate Control (TGC) function was used. Rather than a uniform TGC value, we apply multiple custom-defined TGC values at different distances along the beam in order to measure flow velocities as close as possible to the transducer surface. Even so, a small region close to the transducer surface (usually 5-15 mm) is unmeasurable due to ringing in the transducer 
tip. When the current is turned on for EVF measurements, the signal becomes noticeably noisier due to interference from induced magnetic fields. However, high quality data are still possible.

Finally, for both thermocouples and UDV probes, we hope to capture all the relevant spectral information, requiring that time between samples be shorter than the typical flow timescales. For convection, the buoyancy turnover timescale can be estimated from balancing buoyant and inertial terms in the Navier-Stokes equation: $t_\mathit{ff} = \left(H / \alpha g \Delta T\right)^{1/2}$ \citep{Glazier99}. For electrovortex flow, the velocity scaling from \cite{Herreman19} can be rearranged to give $t_\mathrm{EVF} \sim H/S$ for $S \lesssim 10^3$ and $t_\mathrm{EVF} \sim H/S^{1/2}$ for $S \gtrsim 10^5$. For our most vigorous achievable flows, we estimate a minimum timescale of 1.2 seconds. Both thermal and velocity measurements are acquired at multiple samples per second.

\begin{figure}
\centering
\includegraphics[width=1\linewidth]{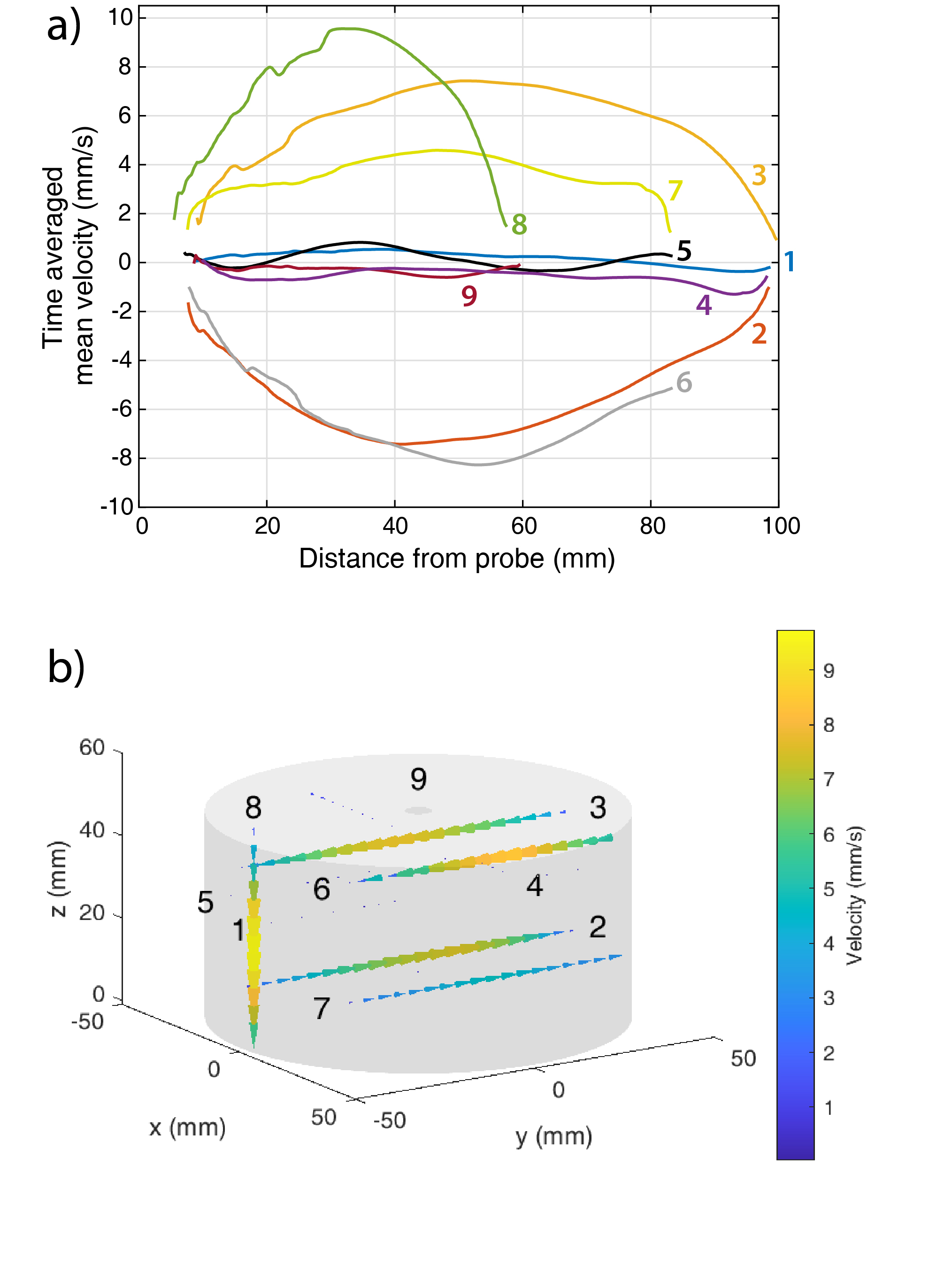}
\caption{\label{F:mov} a) Time-averaged UDV measurements from multiple probes in a convection case at $Ra=3.7\times10^{5}$ and $\Gamma = 1.7$. Colors correspond to the probe positions denoted in Fig.~\ref{F:probes}. b) a frame of a movie produced from the same data (Replace with time-averaged version instead). Velocity vectors are represented by cones pointing in the direction of flow. Cone size and color represent velocity magnitude. Cropping, smoothing, and interpolation in time and space eliminate spurious vectors and noise. See Supplementary Materials for a time-dependent movie in the style of panel b.}
\end{figure}

\section{Typical measurements} \label{sec:data}

Experimental measurements are shown in Figure \ref{F:data}. Noisy regions corresponding to reflections from the back wall of the vessel have been truncated. Measurements from each UDV probe are represented as a Dopplergram in panel a), where velocities along the beam are plotted versus time and space. Panels b) and c) are time- and space-averaged data from the Dopplergram, respectively, and panel d) is the FFT spectrum, where a peak likely associated with the recently characterized jump-rope vortex appears \citep{Vogt18}. In panels a) and b), gaps in the velocity observed near the probe surface are an artifact of the probe tip ringing effect.

Raw UDV datasets as in Figure \ref{F:data} do not lend themselves to easy visualization, and their relationship with the actual flow field may not be obvious. In Figure \ref{F:mov}, we apply a novel post-processing tool to address this issue: time-dependent velocity data from each probe are mapped to their physical position on an analogue to the apparatus and recorded in a movie. Figure \ref{F:mov} shows the time-averaged velocity field; see Supplementary Materials for an animation. We apply a spatial moving average of 30 bins (7 mm) to reduce noise and account for velocity gaps. 

\begin{figure}
\centering
\includegraphics[width=0.9\linewidth]{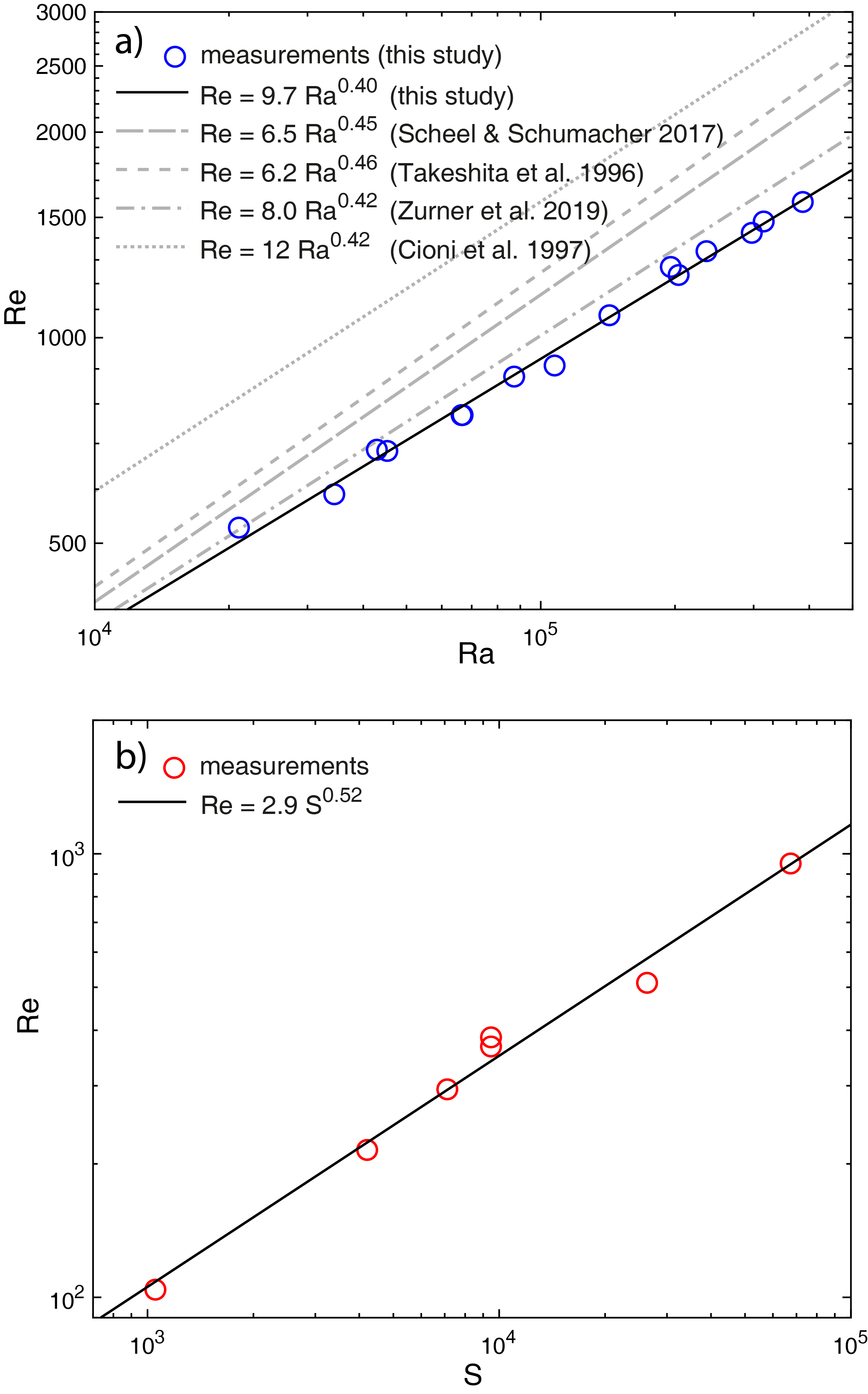}
\caption{\label{F:rera} a) Reynolds number $Re$ plotted versus Rayleigh number $Ra$ for a set of convection cases in the $\Gamma = 2$ setup. Results from \cite{Takeshita96,Cioni97,Scheel17,Zurner19} are also included for comparison. The best-fit scaling, $Re \propto Ra^{0.40}$, matches the exponent 0.4 predicted by theory in Eq.~\ref{eq:rera} \citep{Ahlers09}. b) Reynolds number $Re$ plotted versus EVF parameter $S$ for a set of EVF cases in the $\Gamma = 2$ setup. The best-fit scaling $Re \propto S^{0.52}$ matches closely with the exponent $0.5$ predicted by theory \citep{Bojarevics89}.
}
\end{figure}
%

We examined how $Re$ scales with $Ra$ in order to validate that the convective flows we drive have their expected properties, as shown in Figure \ref{F:rera}a. We performed a series of convection experiments, with no current, by setting adverse temperature gradients of $1-20~^\circ$C between the thermal bath and chiller. To obtain a precise value of $\Delta T$ for calculating $Ra$ via Eq.~\ref{eq:Ra}, we averaged the temperature difference between top and bottom thermocouples, over both space and time. To derive a $Re$ value from velocities, we first observed from Figure \ref{F:mov} that a single large scale circulation (LSC) appears to fill the container. We therefore adopted the method of \cite{Zurner19} to extract a characteristic LSC velocity from UDV probes: the magnitude of the velocity at $\left(x,y,z\right) = \left(0,0,0.75H\right)$ is $U_\mathrm{LSC} = \langle\left(u_3^2+u_4^2\right)^{1/2}\rangle$, where $u_3$ and $u_4$ are the velocities measured by probes 3 and 4, respectively, and $\langle \rangle$ indicates time averaging. Unlike \cite{Zurner19}, we did not average this velocity near the top with a velocity near the bottom, since there is only one probe beam intersecting $\left(x,y,z\right) = \left(0,0,0.25H\right)$. Figure \ref{F:rera}a shows that our data are well-fit by a $Re$--$Ra$ power law with exponent 0.40, in close agreement with the theoretically predicted 0.4, and also in agreement with scaling results from previous convection studies.

We validated EVF results through a similar procedure, as shown in Figure \ref{F:rera}b. We set a small stabilizing temperature gradient $\Delta T = -0.5~^\circ$C and varied the current from 5 to 40 A. We calculated $S$ following Eq.~\ref{eq:S}. To estimate $Re$, we modified the methodology of \cite{Herreman19}: the characteristic EVF velocity was chosen by finding the peak velocity out of all UDV measurements at each time step, and then averaging this value over time, i.e. $U_\mathrm{EVF} = \langle u_\mathrm{max} \rangle$. We find that results bear close agreement to the predicted $Re$--$S$ power law exponent of 0.5. These fits to theoretical predictions and literature values suggest that our thermal controls, current controls, and diagnostics are robust.

\section{Conclusion}
\label{sec:conclusion}

Liquid metal batteries are an evolving technology. Since it is unclear which geometries and materials will ultimately have industrial relevance, our understanding of flows in LMBs must therefore be general enough to accommodate these possibilities. Our new experimental apparatus addresses this problem by isolating two primary flow drivers --- thermal gradients and electrovortex flow --- in order to predict their effects and interactions over a range of governing parameters scalable to LMBs. Compared to a full LMB prototype, we can more clearly pinpoint the causes for each observed flow behavior. This is aided by treating a simplified cylindrical geometry, looking at each layer independently, and conducting experiments near room temperature.

To produce well-characterized flows, we ensure that the container materials provide the correct boundary conditions and that the control systems provide precise temperatures and currents. We validate these systems with simplified theoretical models and experimental tests. To adequately quantify the flows, we inform the placement of temperature and velocity probes with predictions from the literature. Using temperature and velocity measurements, we can calculate governing parameter values and partially reconstruct the flow field. We validate these measurements through comparison to previous results in the literature.

Future studies in our setup will involve surveys of convection, EVF, combined convection and EVF, and combined stable stratification and EVF. In combined surveys, we will check existing predictions for the transition between EVF-dominated and thermal-dominated flows \citep{Davidson01,Ashour18}. Our study of EVF combined with stable thermal stratification will be, we believe, the first of its kind.

While our simplified model has many advantages, care must also be taken when applying our results to LMBs. Thermal gradients could also manifest in other patterns, such as with horizontal components \citep{Shishkina16}. The Lorentz forces driving EVF occur in close proximity to the narrower electrode, meaning geometrical or material differences in that region could have a significant effect on the resultant flow. 
Finally, convection and EVF must be weighed against the multitude of other possible flows. As mentioned in section \ref{sec:motiv}, solutal convection can dominate thermal effects. Fluid-fluid interfaces can serve as a source for many flows: for example, tilted interface surfaces between fluids of different conductivity lead to the metal pad roll instability \citep{Sele77,Zikanov15,Horstmann18}; variations in surface tension along interfaces lead to Marangoni flow \citep{Kollner17}. The relative strengths of each effect are not fully characterized, but \cite{Kelley18} contains an extensive discussion about this topic.

Advancing the study of LMBs requires distinguishing the contributions of myriad flows in a controlled setting.
Our novel setup provides an important step toward understanding the fluid behaviors in LMBs and thereby moving them ever closer to commercial viability.

\section{Appendix: Theoretical model for heat transfer inside heat exchanger plate}

We consider a two-dimensional radial cross-section of the copper plate placed in an $(x,z)$ Cartesian coordinate system with $z$ pointing upwards, $x$ along the diameter of the plate, and the origin at the midpoint of the plate's bottom. Since we are interested in the steady-state temperature distribution of the plate, the governing equation is the Laplace equation:

\begin{equation}
\label{eq:T eqn}
\nabla^{2}T(x,z) = 0.
\end{equation}

The side walls of the plate are in contact with a heat insulator, and we assume that no heat escapes through the sides, i.e., the heat flux through the side walls is zero, which is asserted as the Neumann boundary condition:

\begin{equation}
\label{eq:side BC}
\frac{\partial T}{\partial x}(x = -\frac{D}{2};\frac{D}{2},z) = 0.
\end{equation}

At the plate-gallium interface heating (or cooling) convection exists between the plate boundary and the liquid gallium, and the heat transfer through the boundary will be proportional to the temperature difference between the gallium and the boundary. We use Newton's law of cooling and find the Robin boundary condition: 

\begin{equation}
\label{eq:bottom BC}
\frac{\partial T}{\partial z}(x,z=0) = \frac{-Bi}{h_\mathrm{plate}} (T(x,z=0) - T_{Ga}),
\end{equation}
where $T_{Ga}$ is the gallium's mean temperature.

The coil is in contact with the last plate boundary and we represented its thermal effect as an evenly spaced distribution of hot and cold spots added to a set temperature, which can be stated as the Dirichlet boundary condition:

\begin{equation}
\label{eq:top BC}
T(x,z=h_\mathrm{plate})  = T_\mathrm{set} + \delta T \cos\left(\frac{2\pi x }{\lambda}\right) ,
\end{equation}

where $T_\mathrm{set}$ is the set temperature of the plate, $\delta T$ is the temperature difference between a hot spot and a cold spot, and $\lambda$ is the distance between two hot or two cold spots. 

We solve Eq.~\ref{eq:T eqn} with boundary conditions given by Eqs.~\ref{eq:top BC}, \ref{eq:side BC}, and \ref{eq:bottom BC} numerically using MATLAB PDEModel with a mesh of $\sim$4800 two-dimensional quadratic triangular elements. We use the parameters $Bi = 0.005$, $\lambda = 2.5~cm$, $D = 10~cm$, $T_{Ga} = 43~^{\circ} C$, and a range of plate thicknesses $< 0.6~cm$ with $T_\mathrm{set}$ values equivalent to the ones used in experiment. We set $\delta T$ to 5\% of the thermal gradient applied between top and bottom $\Delta T$. We quantify the temperature variation at the boundary in contact with gallium as the standard deviation of the temperature there, reported as a percentage of $\Delta T$. 
\begin{figure}
\centering
\includegraphics[width=1\linewidth]{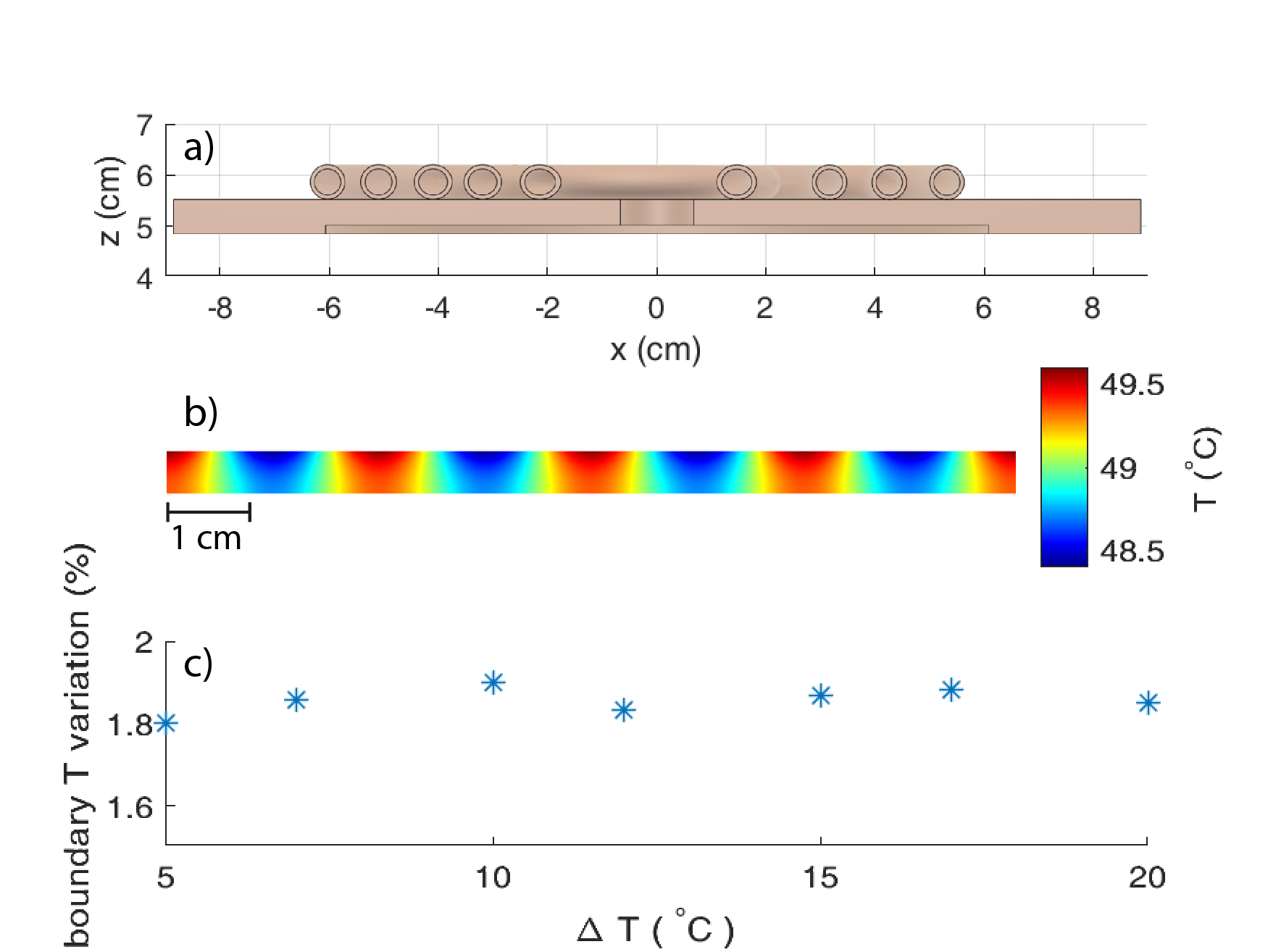}
\caption{\label{F:coilmodel} a) Radial cross-section of the copper plate attached to heat exchanger. b) Steady state temperature distribution inside the plate for $\Delta T = 12$. c) The temperature variation at the plate-gallium boundary for different $\Delta T$ values where the plate thickness is $0.5$~cm. As shown, the variation is $< 2\%$ for all relevant $\Delta T$ cases.}
\end{figure}      

\begin{acknowledgements}
The authors wish to thank Mike Pomerantz and Jim Alkins for design consultation and construction of the setup. The authors also thank Gerrit M.\ Horstmann for productive discussions and the idea of writing this paper, as well as Norbert Weber and Tobias Vogt for fruitful scientific discussions. We thank Alex M.\ Grannan for inspiring the 2D model described in the Appendix, which was built off of his PhD work. 
This work was supported by the National Science Foundation under award number CBET-1552182.
\end{acknowledgements}

%
\section*{Conflict of interest}

The authors declare that they have no conflict of interest.

\bibliographystyle{spbasic}      
\bibliography{Bib_list}   

\end{document}